\documentstyle[epsf,times]{aipproc}

\begin{document}
\title{Determination of ferroelectric compositional phase transition
using novel virtual crystal approach}

\author{Nicholas J. Ramer$^*$ and Andrew M. Rappe$^{\S}$}
\address{$^*$Department of Chemistry, Long Island University -
C. W. Post Campus, Brookville, NY 11548\\ $^\S$Department of
Chemistry and Laboratory for Research on the Structure of Matter,\\
University of Pennsylvania, Philadelphia, PA 19104}

\maketitle

\begin{abstract} 
We employ a new method for studying compositionally disordered
ferroelectric oxides.  This method is based on the virtual crystal
approximation (VCA), in which two or more component potentials are
averaged into a composite atomic potential.  In our method, we
construct a virtual atom with the correctly averaged atomic size and
atomic eigenvalues.  We have used our new method to study the
composition dependent phase transition in Pb(Zr$_{1-x}$Ti$_x$)O$_3$
lying between $x=0.5$ and $x=0.4$.  We correctly predict the
experimentally determined phase transition from the tetragonal phase
to a low-temperature rhombohedral phase between these two
compositions.

\end{abstract}

\section*{Introduction}

Our recent work~\cite{njr:Williamsburg99} has demonstrated the utility of
the virtual crystal approximation (VCA)~\cite{njr:VCA1,njr:VCA2} to study
compositionally disordered materials.  Following our initial VCA study
of a stress-induced phase transition in Pb(Zr$_{1-x}$Ti$_x$)O$_3$, we
and other authors have extended the VCA to studies of compositional
phase transitions~\cite{njr:PRBVCA}, temperature-dependent phase
transitions~\cite{njr:DHVVCA}, and dynamical properties.~\cite{njr:Rabe} Prior
to this the VCA has had some success in providing qualitative, and in
some instances quantitative agreement with large-scale solid-state
calculations.

Previous implementations of the VCA have focused on the averaging of
two or more component potentials in separable form at the solid-state
level.~\cite{njr:Papa,njr:Slavenburg} By averaging the potentials, it is
difficult to assess the quality of the resulting VCA potential.  In
some studies, these traditional VCA potentials have yielded unphysical
results.~\cite{njr:Chen,njr:Bellaiche} Speculation has been made in the
literature that it is the inability of these traditional VCA
potentials to capture the important differences in chemical bonding
and ionicity accurately that has given the poor results.~\cite{njr:Chen}

\section*{Methodology}

Due to the inconsistencies in the literature regarding the quality of
VCA results, we have recently formulated a new approach for the
construction of VCA potentials.  Our method represents a simple and
intuitive way to construct VCA potentials and assess their quality. It
is a severe departure from the more traditional methods for VCA
potential construction.  We have previously presented a thorough
description of our method for the construction of VCA
atoms.~\cite{njr:Williamsburg99,njr:PRBVCA} For brevity, we will only present
here the salient points of our method.

\begin{itemize}
\item{Our method averages the component atoms at the all-electron
level, such that all-electron eigenvalues, potentials, wave functions
and charge density are all computed for the virtual atom.  Since the
solution of the all-electron atom is common to all pseudopotential
procedures, our method is general enough to be used in all types of
pseudopotential construction.

Operationally we find the all-electron wave functions and eigenvalues
for the atoms we want to average.  We then determine the properly
averaged eigenvalues for the VCA atom valence states from the
eigenvalues of the all-electron atoms.  We then construct a bare
Coulombic potential from the properly averaged bare nuclear potentials
of the component atoms and a properly averaged core charge density.
In doing so, we have insured that the resulting VCA potential will
have the correct averaged size.  Using this averaged nuclear potential
and frozen-core charge density, we construct new self-consistent wave
functions for the VCA atom valence states.  We require that these new
wave functions solve the Kohn-Sham equations in the valence region and
give the properly averaged eigenvalues.}

\item{From this new set of potentials and wave functions, we construct
an optimized pseudopotential.~\cite{njr:RappePS} This method enforces the
exact agreement of the VCA eigenvalues with the averaged all-electron
eigenvalues for the reference state {\em only}.  For any other electronic
configuration, we have made no requirements on the quality of the
potential.  Recently, we have formulated the designed nonlocal
pseudopotential method,~\cite{njr:RRPS} in which we exploit the inherent
arbitrary separation of the local potential and semilocal correction
terms required for the Kleinman-Bylander separable form.~\cite{njr:KB} By
adjusting the form of the local potential and therefore the correction
terms, we may dramatically improve the transferability of the
potential without affecting the exact agreement at the reference
configuration.  This transferability improvement over a wide range of
electronic configurations insures correct ionicity and polarizability
of the virtual atom.}

\end{itemize}

It is our assertion that the preservation of the proper averaged size
of the VCA potential and a high level of transferability will provide
accurate VCA atoms which yield excellent agreement with superlattice
calculations.  We have previously demonstrated the effectiveness of
our new VCA approach.~\cite{njr:Williamsburg99,njr:PRBVCA} The remainder of
this paper will focus on the construction and use of two VCA
potentials to study the compositional phase transition in
Pb(Zr$_{1-x}$Ti$_{x}$)O$_3$.
  
\section*{Results and Discussion}

\begin{table}[b]
\caption{Construction parameters for the Zr$_{1-x}$Ti$_x$ virtual
crystal (VCA) designed nonlocal pseudopotentials.  The VCA potentials
were generated with the methods described in references
\protect\onlinecite{njr:RappePS},\protect\onlinecite{njr:RRPS} and the method
described in text.  Core radii ($r_c$) are in atomic units, $q_c$
parameters are in Ry$^{1/2}$, step widths are in atomic units and step
heights are in Ry.}

\begin{tabular}{lcdcd}
    &       &       &Step &Step   \\
Atom&$r_{c}$&$q_{c}$&Range&Height \\ \hline
    &             &       &         &       \\
VCA ($s^{2}p^{6}d^{0}$)&1.38,1.51,1.40&7.07&0.00--0.56 &
1.18   \\ 
$x=0.5$&                                 &              &0.56--0.79 &
1.34   \\ 
    &             &       &         &       \\
VCA ($s^{2}p^{6}d^{0}$)&1.27,1.38,1.61&7.07&0.00--1.25 &
10.32   \\
$x=0.4$&             &       &      &       \\
    &             &       &         &       \\
\end{tabular}
\end{table}

\begin{table}[ht!]
\caption{Configuration testing for Zr$_{1-x}$Ti$_x$ virtual crystal
(VCA) atoms generated with the method described in text.  Averaged
eigenvalues and total energy differences are given for Zr and Ti
pseudopotentials (PS) for two values of $x$.  The PS construction
parameters have been presented
elsewhere.\protect\onlinecite{njr:Williamsburg99,njr:PRBVCA} Errors are
computed as a difference between VCA and averaged PS results.  All
energies are in Ry.}

\begin{tabular}{crr|rr}
 &\multicolumn{2}{c|}{$x=0.5$}&\multicolumn{2}{c}{$x=0.4$}\\
\cline{2-3} \cline{4-5}
&\multicolumn{1}{c}{PS}&\multicolumn{1}{c|}{VCA}&\multicolumn{1}{c}{PS}&\multicolumn{1}{c}{VCA}\\
State&\multicolumn{1}{c}{Energy}&\multicolumn{1}{c|}{Error}&\multicolumn{1}{c}{Energy}&\multicolumn{1}{c}{Error}\\ \hline
 & & & & \\ 
$s^2$&-7.5701& 0.0000&-7.4266& 0.0000\\ 
$p^6$&-5.9530& 0.0000&-5.8303& 0.0000\\
$s^0$&-2.5581&-0.0038&-2.5280&-0.0029\\
$d^0$&-3.4488& 0.0000&-3.3629& 0.0000\\ 
$\Delta E_{\rm{tot}}$& 0.0000& 0.0000& 0.0000& 0.0000\\
 & & & & \\ 
$s^2$&-6.7808& 0.0016&-6.6490& 0.0019\\
$p^6$&-5.1720& 0.0015&-5.0611& 0.0017\\
$s^1$&-2.0304&-0.0009&-2.0061&-0.0013\\
$d^0$&-2.7034& 0.0041&-2.6316& 0.0050\\ 
$\Delta E_{\rm{tot}}$&-2.3067&-0.0019&-2.2790&-0.0014\\ 
 & & & & \\ 
$s^2$&-6.3652&-0.0031&-6.2570&-0.0027\\
$p^6$&-4.7744&-0.0049&-4.6868&-0.0038\\
$s^0$&-1.8399& 0.0000&-1.8234& 0.0004\\
$d^1$&-2.3571&-0.0033&-2.3087&-0.0047\\ 
$\Delta E_{\rm{tot}}$&-2.8936&-0.0012&-2.8270&-0.0007\\
 & & & & \\ 
$s^2$&-5.6586&-0.0029&-5.5599&-0.0023\\
$p^6$&-4.0734&-0.0047&-3.9953&-0.0040\\
$s^1$&-1.3453& 0.0011&-1.3327& 0.0019\\
$d^1$&-1.6889&-0.0018&-1.6517&-0.0010\\ 
$\Delta E_{\rm{tot}}$&-4.4923&-0.0004&-4.4107& 0.0005\\
 & & & & \\
$s^2$&-4.1730&-0.0077&-4.1060&-0.0061\\
$p^6$&-2.6064&-0.0106&-2.5597&-0.0091\\
$s^2$&-0.3293& 0.0018&-0.3285& 0.0026\\
$d^2$&-0.3239&-0.0072&-0.3193&-0.0058\\ 
$\Delta E_{\rm{tot}}$&-6.2696&-0.0037&-6.1628&-0.0022\\ 
 & & & & \\
\end{tabular}
\end{table}

We have applied our new VCA method to the Zr$_{1-x}$Ti$_{x}$ virtual
atom for $x=0.5$ and $x=0.4$.  All atomic energy calculations were
done with the local density approximation and optimized~\cite{njr:RappePS}
and designed nonlocal pseudopotential~\cite{njr:RRPS} methods were
employed.  The generation parameters for the two VCA potentials are
included in Table 1.  For all atoms, semi-core states were included as
valence.  It is important to note that although we have included
multiple $s$-channel states, only one $s$ nonlocal projector is used.
For both VCA atoms, we have used the $s$-potential with the addition
of one or two square-step potentials as the local potential.

The transferability testing results for the two VCA potentials are
contained in Table 2.  From the table, it is evident that our new VCA
pseudopotentials are highly transferable at either value of $x$.

We use both VCA potentials in solid-state calculations for three
structural phases of Pb(Zr$_{1-x}$Ti$_{x}$)O$_3$.  The electronic wave
functions are expanded in a plane-wave basis using a cutoff energy of
50 Ry.  We included the 5$d$ shell as valence for the Pb atom as well
as including scalar relativistic effects.  Brillouin zone integrations
are approximated accurately as sums on a 4$\times$4$\times$4
Monkhorst-Pack $k$-point mesh.~\cite{njr:MNKPCK}

We have included three structurally distinct phases of
Pb(Zr$_{1-x}$Ti$_{x}$)O$_3$ in the current studies of compositions
between $x=0.5$ and $x=0.4$: a tetragonal Ti-rich $P4mm$ and two
rhombohedral Zr-rich ($R3c$ and $R3m$) phases.~\cite{njr:Jaffe,njr:PZTJaffe}
Around 400-500K, the rhombohedral region exhibits a boundary between
$R3c$ (low-temperature) and $R3m$ (high-temperature) phases, which
depends weakly on composition.~\cite{njr:Michel,njr:Glazer} The $R3c$ phase
shows complex oxygen octahedral tilting, which doubles the primitive
unit cell to ten atoms.~\cite{njr:Clarke}

We have completed full electronic and structural relaxations for
five-atom unit cells for the tetragonal and high-temperature
rhombohedral phases.  Due to the complex oxygen octahedral
distortions, we have used a 10-atom unit cell for the low-temperature
rhombohedral phase.  For all rhombohedral calculations, we have
neglected the small shear relaxations.

In Figure 1, we show the equations of state for the three phases of
Pb(Zr$_{1-x}$Ti$_{x}$)O$_3$ at $x=0.5$.  We find the tetragonal phase
is the ground-state structure for this composition lying approximately
0.05 eV lower in energy than both rhombohedral phases.  This is correct
energy ordering found experimentally.~\cite{njr:Jaffe,njr:PZTJaffe}

\begin{figure*}[ht!]
\vspace{-1.6in}\epsfxsize=3.5in
\centerline{\hspace{-1.2in}\epsfbox[75 335 540 730]{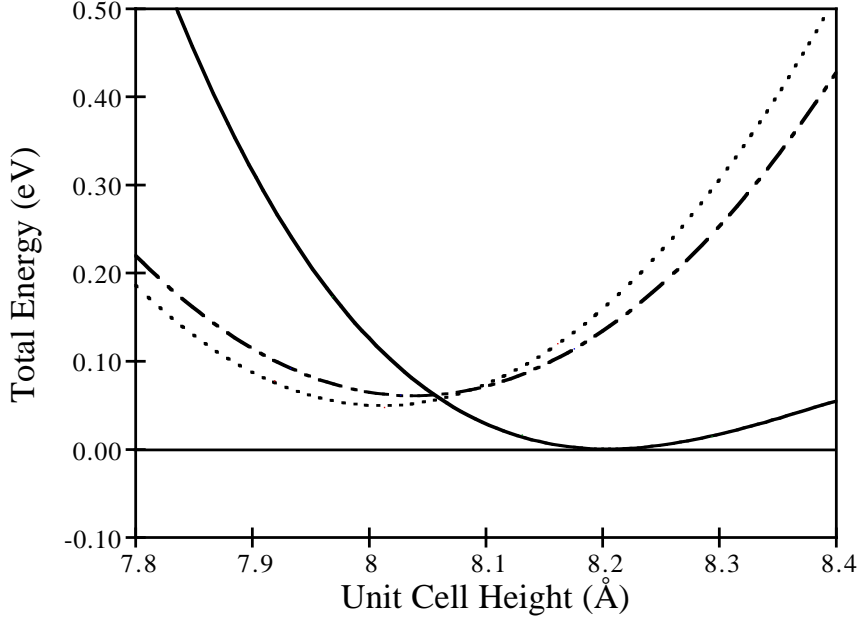}}
\vspace{2.5in}\caption{Equations of state for the tetragonal (solid
line), high-temperature rhombohedral (dotted-dashed line) phase and
low-temperature rhombohedral (dotted line) phases of
Pb(Zr$_{1-x}$Ti$_{x}$)O$_3$ for $x=0.5$.  The heights and energies are
for a 40-atom unit cell.}
\end{figure*}

In Figure 2, we show the equations of state for the three phases of
Pb(Zr$_{1-x}$Ti$_{x}$)O$_3$ at $x=0.4$.  We find both rhombohedral
phases have shifted down in energy relative to the tetragonal phase.
In fact, the low-temperature rhombohedral phase has shifted by
approximately 0.09 eV in energy and therefore becomes the ground-state
structure for this composition of Pb(Zr$_{1-x}$Ti$_{x}$)O$_3$.  This
result is in direct agreement with the experimental
findings~\cite{njr:Jaffe,njr:PZTJaffe} and demonstrates the ability of our new
VCA method to locate and predict compositional phase transitions.

\begin{figure*}[ht!]
\vspace{-1.6in}\epsfxsize=3.5in
\centerline{\hspace{-1.2in}\epsfbox[75 335 540 730]{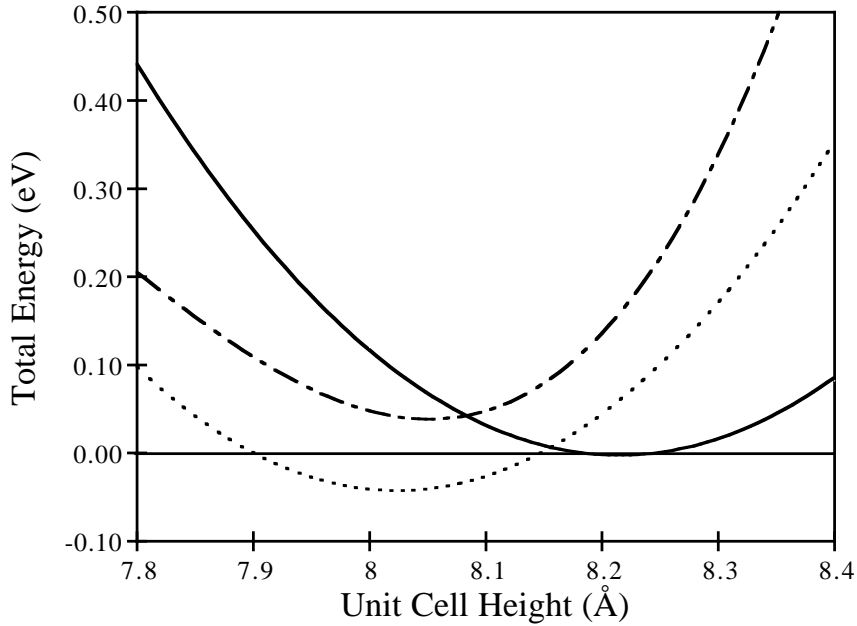}}
\vspace{2.5in}\caption{Equations of state for the tetragonal (solid
line), high-temperature rhombohedral (dotted-dashed line) phase and
low-temperature rhombohedral (dotted line) phases of
Pb(Zr$_{1-x}$Ti$_{x}$)O$_3$ for $x=0.4$.  The heights and energies are
for a 40-atom unit cell.}
\end{figure*}

\section*{Conclusions}

In this paper, we have applied our new method for constructing a
virtual crystal pseudopotential to the Zr and Ti atoms.  Potentials
constructed with our method not only possess the properly averaged
atomic size but also a high level of accuracy in describing energetic
differences due to changes in both ionicity and polarizability.  This
method is based on averaging all-electron information and therefore
makes the method applicable to all types of pseudopotential
construction algorithms.  We have used our new method to construct two
Zr$_{1-x}$Ti$_{x}$ VCA potentials at $x=0.5$ and $x=0.4$.  With these
potentials, we have studied three structural phase of
Pb(Zr$_{1-x}$Ti$_{x}$)O$_3$, finding that the tetragonal phase
is the ground-state structure for Pb(Zr$_{1-x}$Ti$_{x}$)O$_3$ at
$x=0.5$.  When moving to $x=0.4$, we find that the low-temperature
rhombohedral phase becomes the ground-state.  This finding is in
excellent agreement with experiment and represents the first $ab$
$initio$ determination of a compositionally-dependent phase transition
in a ferroelectric oxide.

Recently yet another phase (monoclinic $Cm$) has been experimentally
found at low temperatures for compositions between $x=0.45$ and
$x=0.5$.~\cite{njr:Cross} We plan to extend the current research by
studying the monoclinic phases with our VCA atoms.

\section*{Acknowledgments}

The authors would like to thank Ilya Grinberg for his help with
programming the virtual crystal method.  This work was supported by
NSF grant DMR 97-02514 and the Laboratory for Research on the
Structure of Matter at the University of Pennsylvania.  AMR
acknowledges the Alfred P. Sloan Foundation.  Computational support
was provided by the San Diego Supercomputer Center and the National
Center for Supercomputing Applications.

\end{document}